\documentclass[aps,prb,twocolumn,amsmath,amssymb,superscriptaddress,floatfix,nofootinbib]{revtex4-2}
\usepackage{tabularx}
\usepackage{bm}
\usepackage{euscript}
\usepackage{graphicx}
\usepackage{color}
\usepackage{amsfonts}
\usepackage{amsmath}
\usepackage{physics}
\usepackage{exscale}
\usepackage{amsbsy}
\usepackage{textcomp}
\usepackage{hyperref}
\usepackage{subcaption}
\usepackage{tikz}
\usepackage{amsmath, amssymb}
\usepackage{bbold}
\DeclareUnicodeCharacter{2215}{/}

\def\flippablevert{\tikz[baseline=.1ex]{
		\fill (0,0) circle (1pt) coordinate (A);
		\fill (0,1.5ex) circle (1pt) coordinate (B);
		\fill (1.5ex,0) circle (1pt) coordinate (C);
		\fill (1.5ex,1.5ex) circle (1pt) coordinate (D);
		\draw (A)--(B);
		\draw (C)--(D);
	}
}

\def\flippablehorz{\tikz[baseline=.1ex]{
		\fill (0,0) circle (1pt) coordinate (A);
		\fill (0,1.5ex) circle (1pt) coordinate (B);
		\fill (1.5ex,0) circle (1pt) coordinate (C);
		\fill (1.5ex,1.5ex) circle (1pt) coordinate (D);
		\draw (A)--(C);
		\draw (B)--(D);
	}
}

\def\tripledimerdown{\tikz[baseline=.1ex]{
		\fill (0,0) circle (1pt) coordinate (A);
		\fill (0,1.5ex) circle (1pt) coordinate (B);
		\fill (1.5ex,0) circle (1pt) coordinate (C);
		\fill (1.5ex,1.5ex) circle (1pt) coordinate (D);
		\draw (A)--(D);
		\draw (B)--(D);
		\draw (C)--(D);
	}
}

\def\flippablehexa{\tikz[baseline=-0.5ex]{
		\fill (1ex,0) circle (1pt) coordinate (A);
		\fill (0.5ex,-0.866ex) circle (1pt) coordinate (B);
		\fill (-0.5ex,-0.866ex) circle (1pt) coordinate (C);
		\fill (-1ex,0) circle (1pt) coordinate (D);
		\fill (-0.5ex,0.866ex) circle (1pt) coordinate (E);
		\fill (0.5ex,0.866ex) circle (1pt) coordinate (F);
		\draw (B)--(C);
		\draw (D)--(E);
		\draw (F)--(A);
	}
}
\def\flippablehexb{\tikz[baseline=-0.5ex]{
		\fill (1ex,0) circle (1pt) coordinate (A);
		\fill (0.5ex,-0.866ex) circle (1pt) coordinate (B);
		\fill (-0.5ex,-0.866ex) circle (1pt) coordinate (C);
		\fill (-1ex,0) circle (1pt) coordinate (D);
		\fill (0.5ex,0.866ex) circle (1pt) coordinate (E);
		\fill (-0.5ex,0.866ex) circle (1pt) coordinate (F);
		\draw (A)--(B);
		\draw (C)--(D);
		\draw (E)--(F);
	}
}

\newcommand{\ZZ}{\mathbb{Z}}

\begin{document}
\title{Quantum orders in the frustrated Ising model on the bathroom tile lattice}

% Reorder list
\author{Sumner N. Hearth}
\affiliation{Department of Physics, Boston University, Boston, Massachusetts 02215, USA}
\author{Siddhardh C. Morampudi }
\affiliation{Center for Theoretical Physics, Massachusetts Institute of Technology, Cambridge, MA 02139, USA}
\author{Chris R. Laumann}
\affiliation{Department of Physics, Boston University, Boston, Massachusetts 02215, USA}

\date{\today}

\begin{abstract}
We determine the zero and finite temperature phase diagram of the fully frustrated quantum Ising model on the bathroom tile (4-8) lattice. 
The phase diagram exhibits a wealth of 2+1d physics, including 
1. classical Coulomb dimer liquids of both square and triangular lattice types;
2. quantum order-by-disorder induced phases breaking $\ZZ_4$, $\ZZ_6$, and $\ZZ_8$ symmetries; 
3. finite temperature Kosterlitz-Thouless (KT) phases floating over the $\ZZ_6$ and $\ZZ_8$ orders; 
and, 4. staircases of (in)-commensurate symmetry breaking phases at intermediate coupling.
We establish this elaborate phase diagram using a combination of dimer model mapping, perturbation theory, Landau analysis and Stochastic Series Expansion Quantum Monte Carlo (QMC-SSE). 
Our results provide a baseline for studying frustrated magnetism with D-Wave architecture annealers, where the 4-8 lattice can be embedded naturally without `cloning', reducing the number of competing energy scales.
Simulations with the D-Wave 2000Q demonstrate qualitative agreement with the high temperature portion of the phase diagram, but are unable to access the low temperature phases.
\end{abstract}

\maketitle

\tableofcontents

\section{Introduction}

Programmable quantum annealers have developed significantly in the last decade and now host thousands of interacting qubits realizing an effective transverse field Hamiltonian\cite{Bunyk2014,boothby2020,boothby2021}.
While these developments are primarily motivated by the hope that such quantum annealers will efficiently solve classical optimization problems, there is accumulating evidence that they may be used to probe the statistical physics of frustrated magnets \cite{King_2018,King_2021,harris2018,Kairys2020}.
Decades of work in magnetism has revealed the richness and complexity that frustration imparts on the resulting magnetic phases, and also the  limitations of classical methods for studying them.
This raises the question of whether current or near term quantum annealers can provide new insights into frustrated magnetic models.

\begin{figure}
\centering
    \includegraphics{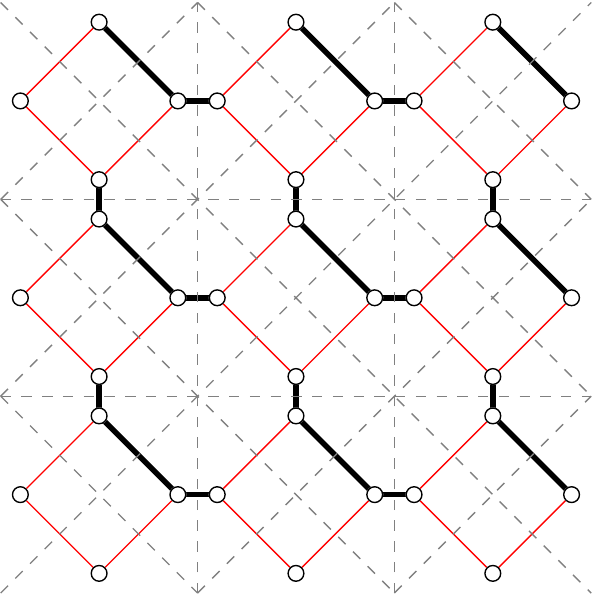}
    \caption{(color online) The direct 4-8 lattice (solid) and the dual Union Jack lattice (dashed). Spins live on circles.  
    In our `gauge' choice, red/thin edges are antiferromagnetic and black/thick are ferromagnetic. 
    In the anisotropic deformation of the model, the thick edges carry coupling $J'$ and the thin have coupling $J$.}
    \label{fig:48embedding}
\end{figure}

This is a delicate undertaking. 
Typically, the large degeneracy of the low energy manifold in frustrated models leads to an extreme sensitivity to perturbations. 
The complexity of embedding such a model into the available architecture thus introduces additional opportunities for systematic errors to creep into the analysis.
Motivated by the underlying architecture of the D-Wave devices -- both the original Chimera and the more recent Pegasus architectures -- we identify the geometrically frustrated transverse field Ising model on the square-octogonal (`4-8') lattice as a natural choice for study as it embeds directly without additional energy scales being introduced by `cloning', where multiple physical qubits are tied together with strong couplings in order to represent a single lattice site (see Fig.~\ref{fig:48embedding}).
Prior work on frustrated magnets with the D-Wave device can in fact be interpreted as studying particular regimes of the 4-8 model where certain bonds have become especially strong\cite{King_2018,King_2021,harris2018,Kairys2020}.
We show, using a variety of theoretical and numerical techniques, that the  larger phase diagram exhibits a cornucopia of classical and quantum phases, capturing a large swathe of statistical physics over the last few decades.  These include two varieties of zero temperature classical Coulomb spin liquid\cite{Henley_2010, Chalker_2017}, adjacent quantum ordered-by-disorder symmetry breaking phases\cite{Villain_Bidaux_Carton_Conte_1980,Andre_Bidaux_Carton_Conte_DeSeze_1979,shender1982anti}, intermediate coupling incommensurate symmetry breaking phases\cite{Bak_1982} and several associated Kosterlitz-Thouless (KT) phases at finite temperature \cite{Kosterlitz_1974, Jose_Kadanoff_Kirkpatrick_Nelson_1977, Isakov_Moessner_2003}.

The results of our case study of the model using the D-Wave 2000Q are qualitatively consistent with our classical analysis in the high temperature regime. 
However, the region of coupling space accessible in the device lies  outside the interesting finite transverse field phases: 
certainly the low temperature ordered phases are entirely inaccessible,
while the finite temperature KT phase is at the margin of accessibility and we see at best qualitative agreement with our computed phase diagram. 
While this is a disappointing result for the current device, the parametric improvements necessary to get into the low temperature phase seem entirely plausible for next generation devices.
We note that the Stochastic Series Expansion Quantum Monte Carlo (QMC-SSE) techniques which we used in our classical study were also inefficient in the low temperature regime. 
This suggests that the D-Wave devices could reliably surpass existing classical techniques in a few generations if these parameters can indeed be improved. 

In the following, we introduce the model more precisely and analyze its symmetries (Sec.~\ref{sec:model_and_symmetries}) and then turn to a tour of its phase diagram using a variety of analytic and numerical techniques, beginning with the classical model at zero transverse field (Sec.~\ref{sec:classicalIsing}), then the finite transverse field model in the isotropic (Sec.~\ref{sec:isotropicmodel}) model and the large (Sec.~\ref{sec:stronglyanisotropicmodel}) and small (Sec.~\ref{sec:weaklyanisotropicmodel}) anisotropy limits.
We connect these with a study of the incommensurate phases at intermediate anisotropy (Sec.~\ref{sec:incommensuratephases}).
We briefly sketch the details of our QMC-SSE implementation and semi-classical updates we introduced to better equilibrate the model in Sec.~\ref{sec:qmc}.
We turn to the results from the D-Wave annealer in Sec.~\ref{sec:annealer} and finally briefly conclude.

\section{Model and Symmetries} % (fold)
\label{sec:model_and_symmetries}

The \emph{isotropic} Frustrated Quantum Ising Model (FQIM) is governed by the transverse field Ising Hamiltonian
\begin{align}
\label{eq:ham_fqim}
    H &= - J \sum_{\langle ij \rangle} s_{ij} \sigma^z_i \sigma^z_j - \Gamma \sum_i \sigma^x_i
\end{align}
where $\langle ij \rangle$ runs over the nearest neighbors of the 4-8 lattice, $\sigma^{x/z}_i$ are the Pauli $x$ and $z$ matrices at site $i$, and $s_{ij} = \pm 1$ determines whether the bond $ij$ is ferro- ($s_{ij} = +1$) or antiferromagnetic ($s_{ij}=-1$). 
The couplings $J$ and $\Gamma$ are positive.
We use indices $i,j$ to indicate general sites on the 4-8 lattice. 
Where needed, we also label sites $i=(\vec{R}, \mu)$ where $\vec{R}$ runs over the sites of a square Bravais lattice (with lattice constant 1) and $\mu$ runs over the 4 sites in the unit cell at the north, east, south and west corners of the elementary square plaquette.

An elementary plaquette $p$ (which can be either a square or an octagon) is classically frustrated if there are an odd number of antiferromagnet bonds around its boundary; mathematically, $p$ is frustrated if \mbox{$F_p=\prod_{\langle ij \rangle \in \partial p} s_{ij} = -1$}.
The fully frustrated model has $F_p = -1$ for every plaquette. 
Without loss of generality, we choose a `gauge' $s_{ij}$ where the antiferromagnet bonds are arranged on the thick bonds of Fig.~\ref{fig:48embedding} -- ie. the horizontal and vertical bonds between the squares and the northeast edge of each square. 
It is useful to define a `gauge' transformation by $\chi_i \in \{ \pm 1 \}$ to flip the $z$-component of each spin $i$ for which $\chi_i = -1$; that is,
\begin{align}
    G(\{\chi_i\})= \prod_i (\sigma^x_i)^{\frac{1-\chi_i}{2}}
\end{align}
Except for the global Ising flip,
\begin{align}
    G_2 = G(\{\chi_i = -1\}) = \prod_i \sigma^x_i
\end{align}
such transformations are manifestly not symmetries of the Ising Hamiltonian $H$ as they send $H(\{s_{ij}\})$ to the distinct Hamiltonian, $H(\{\chi_i s_{ij} \chi_j \})$.
On the other hand, the frustrations $F_p$ are `gauge'-invariant in that the `gauge'-related Hamiltonians have the same frustrated plaquettes.
We use quotes around the term `gauge' to remind the reader that the FQIM Hamiltonian does not have true local symmetries.

\begin{figure}
\centering
\includegraphics{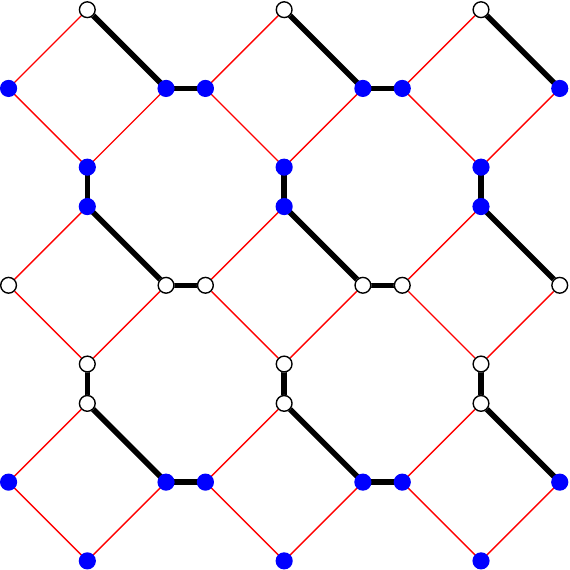}
\caption{The `gauge' transform $G_x$ flips the spins marked in solid blue, mapping the Hamiltonian to another with the same frustrated plaquettes but the choice of frustrated bonds mirrored across the $y$-axis.}
\label{fig:gaugetransform}
\end{figure}

The `gauge' transforms play a useful role in understanding the global symmetry group of the FQIM Hamiltonian. 
The space group of the 4-8 lattice is that of the underlying square Bravais lattice with $\pi/2$ rotation centers at the center of each of the elementary (diagonal) squares. 
This group is generated by lattice translations $T_x$, $T_y$ and the reflections $M_{xy}: (x,y) \to (y,x)$ and $M_x: (x,y) \to (-x,y)$. 
In our `gauge', the Hamiltonian $H$ is clearly invariant under the lattice translations and $M_{xy}$ (see Fig.~\ref{fig:48embedding}), but, the antiferromagnetic bond on the northeast edge of the elementary square is mapped to the northwest edge by $M_x$. 
This can be corrected by an appropriate `gauge'-transformation $G_x = G(\{e^{i(0,\pi)\cdot \vec{R}} (-1)^{\delta_{\mu,1}}\})$ which maps the antiferromagnetic bonds back to the northeast edges (Fig~\ref{fig:gaugetransform}).
Putting this together with the global Ising flip, the global symmetry group $\mathcal{G}$ of $H$ is generated by the operators 
\begin{align}
    \mathcal{G} = \langle G_2, T_x, T_y, M_{xy}, G_x M_x \rangle.
\end{align}
Physically, the isotropic FQIM has the full symmetry of the 4-8 lattice, but the action of the symmetry transformations mixes space and spin degrees of freedom.
This will play an important role in the Landau analysis of the symmetry breaking orders induced by the transverse field.

We further consider an \emph{anisotropic} deformation of the FQIM in which the ferromagnetic bonds (thick bonds in Fig.~\ref{fig:48embedding}) come with magnitude $J'$ while the antiferromagnetic bonds comes with $J$. 
\begin{align}
    H = -J \sum_{\substack{\langle ij \rangle \\s_{ij} = -1}} s_{ij} \sigma^z_i \sigma^z_j - J' \sum_{\substack{\langle ij \rangle \\s_{ij} = +1}} s_{ij} \sigma^z_i \sigma^z_j - \Gamma \sum_i \sigma^x_i
    \label{eq:hfulljprime}
\end{align}
This model has manifestly reduced symmetry -- if a space group transformation moves the northeast bond, it can no longer be fixed by a `gauge'-transformation which only move the signs $s_{ij}$ around. 
Thus, the global symmetry group is reduced to 
\begin{align}
    \mathcal{G}' = \langle G_2, T_x, T_y, M_{xy} \rangle.
\end{align}

Classically, the dual representation of the Ising model\cite{kogut1979} is expressed in terms of the `broken bond' variables,
\begin{align}
    \tau^z_{ij} = s_{ij} \sigma^z_i \sigma^z_j
\end{align}
which are subject to the constraint
\begin{align}
\label{eq:ham_igt_gauss}
   \prod_{\langle i j \rangle \in \partial p} \tau_{ij}^z = \left(\prod_{\langle i j \rangle \in \partial p} s_{ij}\right) \left(\prod_{i \in \partial p} (\sigma_i^z)^2\right) = F_p.
\end{align}
If $\tau^z = -1$, the corresponding bond in $H$ is broken -- i.e., it carries positive energy.
The full quantum Hamiltonian $H$ can be recast as an Ising gauge theory in the $\tau$ variables,
\begin{align}
\label{eq:ham_igt}
    H &= - J \sum_{\langle ij \rangle} \tau^z_{ij} - \Gamma \sum_i \prod_{j \in \partial i } \tau^x_{ij}
\end{align}
where $j \in \partial i$ runs over the sites neighboring $i$ on the lattice. 
Here, $\tau^x_{ij}$ is the Pauli $x$ matrix conjugate to $\tau^z_{ij}$ associated to bond $\langle ij \rangle$.
For simplicity, we have written $H$ for the isotropic case; the anisotropic extension is straightforward.
As usual, the gauge representation, Eq.~\eqref{eq:ham_igt} with constraint Eq.~\eqref{eq:ham_igt_gauss}, is an exact rewriting of the FQIM, Eq.~\eqref{eq:ham_fqim}, up to the global Ising symmetry: 
every valid broken bond configuration $\tau^z$ fixes a $\sigma^z$ configuration up to a  global choice of $\uparrow$ or $\downarrow$. 

In the fully frustrated model $F_p = -1$, so physical configurations must have an odd number of broken bonds around any plaquette.
The classical ground states thus have one broken bond per plaquette and can be profitably reinterpreted as hard-core dimer configurations on the dual union-jack lattice (dashed lines, Fig.~\ref{fig:48embedding}) by drawing each broken bond as a dimer bisecting the direct lattice bond. 
See Fig~\ref{fig:48columnar} for an example such dimer covering corresponding to a `columnar state' of the dimers.
Unlike the essentially exact gauge representation, the hard-core dimer representation is only useful at low temperature and transverse field, where the relevant states are close to the classical ground state manifold.
In this case, however, as much is known about both classical\cite{Kenyon_2003} and quantum dimer models\cite{Rokhsar_Kivelson_1988,Moessner_Raman_2008}, it provides a wealth of guidance.

\begin{figure}
\includegraphics{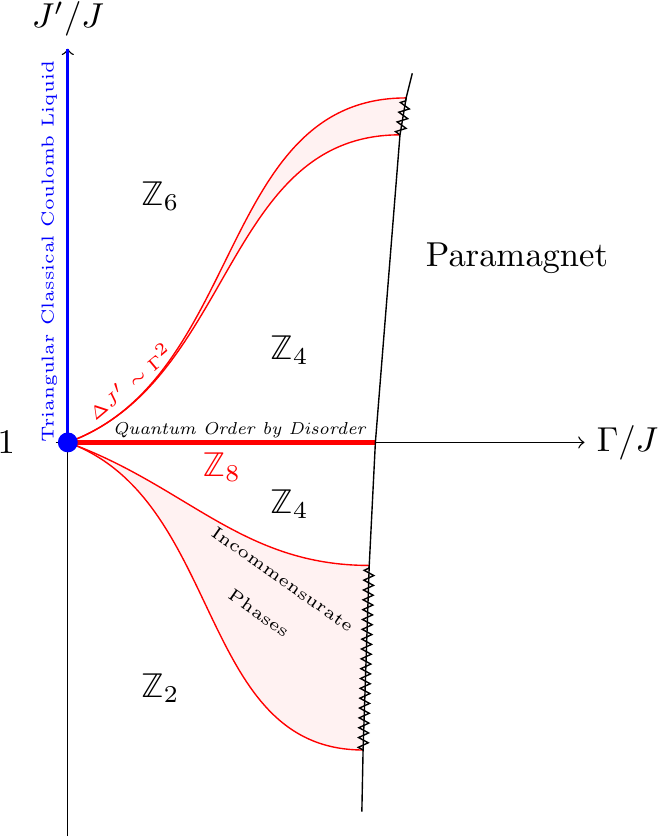}
\caption{The zero temperature phase diagram of the 4-8 fully frustrated Ising model. The blue dot at $J'=J, \Gamma=0$ is a square lattice classical Coulomb liquid. Between the $\ZZ_4$ and $\ZZ_2$ symmetry-breaking phases there is a \emph{staircase} of ordered phases with incommensurate spatial periods. A similar region may exist between $\ZZ_4$ and $\ZZ_6$.}
    \label{fig:phasefull}
\end{figure}

\section{Classical Limit \texorpdfstring{($\Gamma = 0$)}{}}
\label{sec:classicalIsing}

\begin{figure}
    \centering
    \includegraphics{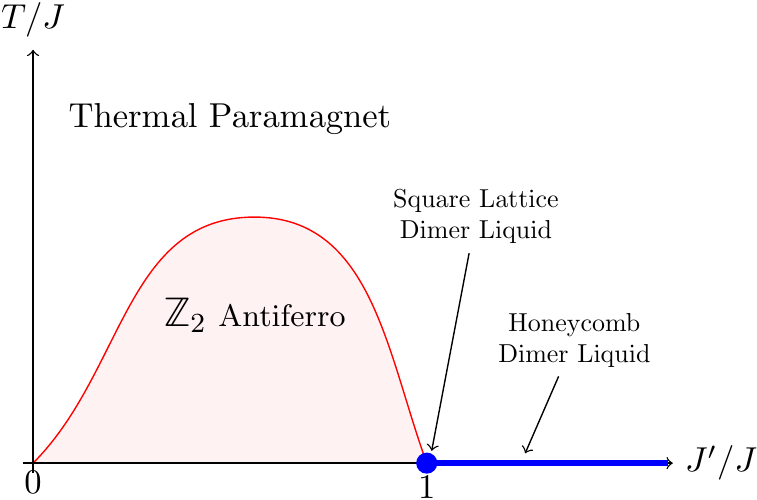}
    \caption{The classical $\Gamma=0$ phase diagram. At $T=0$ there are two regimes: 
    for $J'>0$ the ground states are identical to those of the ZZD Square Lattice\cite{Andre_Bidaux_Carton_Conte_DeSeze_1979} model including a square lattice dimer liquid at $J'=J$ and a honeycomb lattice dimer liquid at $J'>J$; for $J'=0$ the system decouples into disconnected  length 4 Ising chains.}
    \label{fig:zzdphasediagram}
\end{figure}

We start by considering the low temperature phases of the $\Gamma=0$ classical model, see Fig~\ref{fig:zzdphasediagram}.
These can best be understood in the dimer representation on the dual Union-Jack lattice.
This lattice may be viewed as a square lattice oriented at $45^{\circ}$ to the horizontal with additional horizontal and vertical bonds.
All valid dimer coverings of the square lattice are coverings of the Union-Jack lattice; any dimer coverings using the additional horizontal and vertical bonds require additional dimers, and are therefore not ground states\cite{Wu_2006}.
Since the ground state dimer coverings are in 1-to-1 correspondence with the square lattice coverings we recover the well-known residual entropy\cite{Villain_Bidaux_Carton_Conte_1980} of the well studied square lattice Coulomb dimer liquid,

\begin{align}
    \mathcal{S}(T=0) &= 0.583.
    \label{eq:entropyiso}
\end{align}

\begin{figure}
\centering
\includegraphics{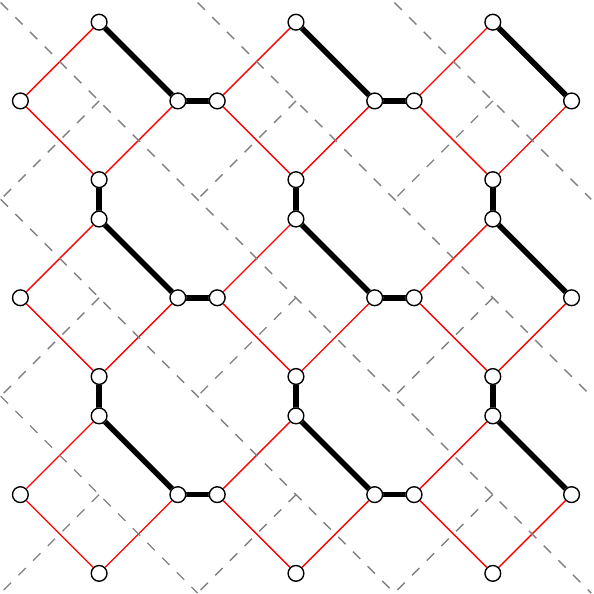}
\caption{At $T=\Gamma=0$ and $J'>J$, the classical ground states are in 2-1 correspondence with a dimer liquid on the dual brick lattice (shown dashed). This is a deformed honeycomb lattice, dual to the triangular lattice.}
\label{fig:48embeddingtriangular}
\end{figure}

The anisotropic deformation strengthening the ferromagnetic bonds by a factor of $J'/J$ bind the spins together and prohibit broken bonds between them at zero temperature (see bold bonds in Fig~\ref{fig:48embedding}).
Removing the appropriate edges from the Union-Jack dimer model produces a brick lattice (Fig~\ref{fig:48embeddingtriangular}) --- a simple deformation of the Honeycomb lattice.
In the classical zero temperature limit we therefore find a honeycomb lattice dimer liquid with the residual entropy\cite{Wannier_1950,Andre_Bidaux_Carton_Conte_DeSeze_1979,Moessner_Sondhi_Chandra_2000},

\begin{align}
    \mathcal{S}'(T=0) &= 0.323.
    \label{eq:entropyaniso}
\end{align}

Weakening the same bonds ($J'/J < 1$) prefers a particular dimer position on each unit cell,
leading to a unique staggered dimer ground state. In the spin language this state breaks the global $\ZZ_2$ symmetry, since dimer states are in a 2-to-1 correspondence with spin configurations.

The $T,J' \ll J$ model, after a gauge transformation, locks spins together within each unit cell and can be mapped to the 2D square lattice ferromagnetic Ising model with bond strength $J'$.
With $J'$ the only accessible energy scale we expect a thermal phase transition to disorder with critical temperature following $T_c \propto J'$.

For $J'/J > 0$ the model has identical zero temperature ground states to those of the square lattice ZZD (\emph{zig-zag domino}) model studied by Andr\'{e} \emph{et al}\cite{Andre_Bidaux_Carton_Conte_DeSeze_1979}.
They predict no direct finite temperature transition and no classical (thermal) order-by-disorder for $J' \ge J$. 
We expect the same for the analogous 4-8 model.

\section{Quantum Model ($\Gamma > 0$)}
\label{sec:symmbreakingphases}

\subsection{Isotropic Model}
\label{sec:isotropicmodel}

\begin{figure}
\centering
\includegraphics{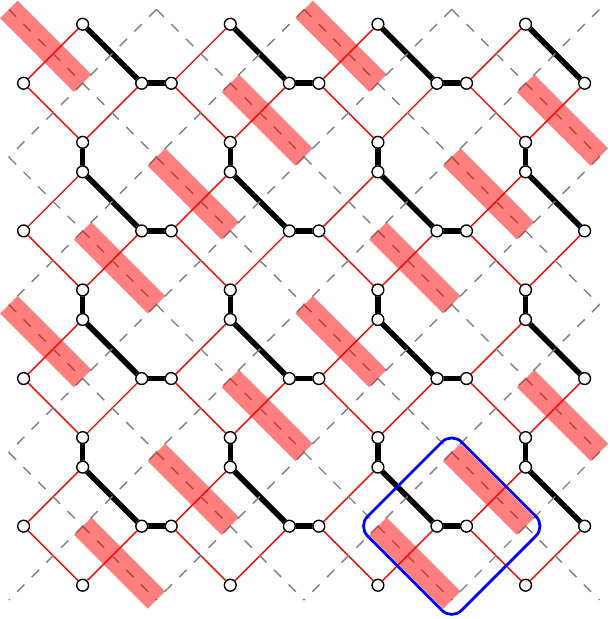}
\caption{An example of a columnar state found at $J' \ge J$. Dimers, which live on the dual lattice, are shown as red rectangles. 
A \emph{flippable plaquette}, such as the one marked with a blue square, lives between each pair of parallel adjacent dimers.}
\label{fig:48columnar}
\end{figure}

To study the ordering effects of a transverse field on the lattice we return to the isotropic case.
A small transverse field lifts the ground state degeneracy of the square lattice dimer liquid, breaking lattice symmetries through \emph{quantum order-by-disorder}\cite{Henley_1989,Moessner_Raman_2008}.
Motivated by the $T=0$ classical mapping to the fully frustrated square lattice we consider symmetry breaking phases which correspond to columnar (Fig~\ref{fig:48columnar}) and plaquette orders of the quantum dimer model.
It is still controversial whether the columnar, plaquette, and a mixed phase is realized on the square lattice.

As in that model, the relevant orders can be organized into a complex order parameter $\phi$ with $\ZZ_8$ clock symmetry.
For the 4-8 lattice, we construct $\phi$ explicitly by considering the  a \emph{single mode approximation}\cite{Moessner_Sondhi_Chandra_2000,feynman_sma,Girvin_2005} (SMA) for magnon excitation in the large $\Gamma$ paramagnetic phase. 

In the infinite $\Gamma$ limit, the ground state is $\ket{\rightarrow \rightarrow ...}$.
Delocalized magnon excitations are created by the operator $\sum a^\pm_\mu e^{\pm i k R} \sigma^z_{R,\mu}$.
For the isotropic model they have minimum energy at $k = (\frac{\pi}{2}, -\frac{\pi}{2})$ with
\begin{align}
    E &= 2\Gamma - (1 + \sqrt{2}) J \label{eq:smaeigensimplified} \\
    a^\pm &= \frac{1}{2}(e^{\pm\frac{3i\pi}{4}}, e^{\pm\frac{i\pi}{2}}, e^{\mp\frac{3i\pi}{4}}, 1).
\end{align}

Eq~(\ref{eq:smaeigensimplified}) predicts a phase transition at $\Gamma_c~=~\frac{1+\sqrt{2}}{2}J$ as the gap closes and magnons condense.
The condensation occurs at finite momentum, breaking lattice symmetries.
Motivated by this pattern, we consider the complex order parameter $\phi$ with momentum $k = (\frac{\pi}{2}, -\frac{\pi}{2})$,

\begin{align}
    \phi &= \sum_{R, \mu} a^+_\mu e^{i k R} \sigma^z_{R,\mu}
    \label{eq:orderparamsymm}
\end{align}

The lattice symmetry group acts on the 2D operator space spanned by $\phi$ and $\phi^\dagger$.
The most salient feature is that $G_x M_x M_{xy}$ acts as a generator of $\ZZ_8$ symmetry, $G_x M_x M_{xy}: \phi \rightarrow e^{-\frac{3 i \pi}{4}} \phi$. Furthermore $T_x G_x M_x: \phi \rightarrow \phi^\dagger$ acts as complex conjugation on the $\phi$ plane.

With this symmetry action in hand, we construct a Landau free energy functional for the condensation transition.

\begin{align}
    \phi &\equiv \vert \phi \vert e^{i\theta} \\
    f &\simeq \sum_{n=1}^{4} c_n \vert \phi \vert^{2n} + g_8 \vert \phi \vert^8 \mbox{cos}(8\theta)
    \label{eq:landauf8}
\end{align}

The first allowed anisotropy for $\phi$ is at eighth order and describes an eight-fold degenerate symmetry breaking ground state.
A value of $g_8 > 0$ favors four columnar states and their global spin flips, whereas $g_8 < 0$ favors plaquette states.
We expect to find columnar states based on the surrounding phases in parameter space, however the analogous  region of the phase diagram on the square lattice is controversial\cite{Wenzel_Coletta_Korshunov_Mila_2012,Yan_Zhou_Syljuasen_Zhang_Yuan_Lou_Chen_2021}.

\begin{figure}
    \includegraphics{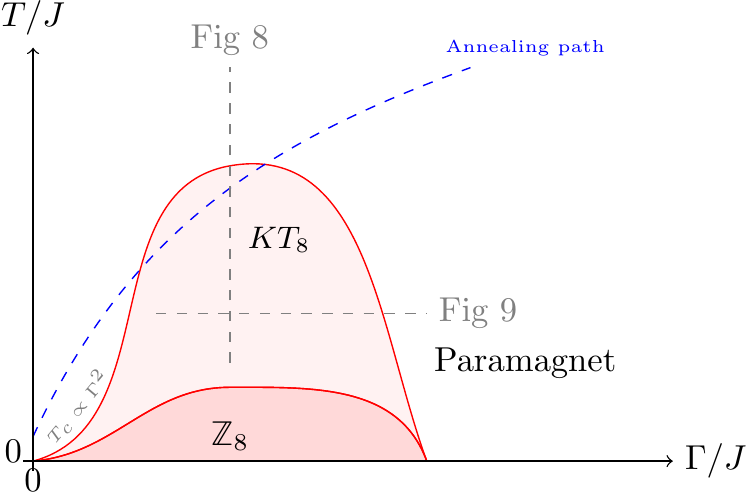}
	\caption{Temperature-transverse field phase diagram of isotropic model ($J'=J$). An example quantum annealing path is shown (see Fig.~\ref{fig:dwavephase}).}
	\label{fig:phaseequal}
\end{figure}

As a discrete symmetry breaking order, the $\ZZ_8$ phase should persist to finite T.
The eight-fold anisotropy is irrelevant, however, at the quantum critical point ($T = 0$, $\Gamma = \Gamma_c$) where we expect a $3D$ XY phase transition to the quantum paramagnet.
Around this point the phase boundaries are controlled by the universal critical exponents of the $3D$ XY transition, producing a quantum critical fan.
The thermal transition is not direct, however, and the thermal critical point splits into a KT critical phase\cite{Kosterlitz_1974,Jose_Kadanoff_Kirkpatrick_Nelson_1977,Isakov_Moessner_2003} with an upper ($T_c^+$) and lower ($T_c^-$) finite temperature transition. 
We expect that the KT phase terminates at $\Gamma = 0$ and $T = 0$ in the dimer liquid.
This can be seen by Andr\'{e} \emph{et al}'s analysis\cite{Andre_Bidaux_Carton_Conte_DeSeze_1979} of the \emph{ZZD} model in which thermal fluctuations simply disorder the square lattice dimer liquid at $T=0$.
The upper critical temperature for the transition from KT to a thermal paramagnet scales with the energy contribution from flippable plaquettes (Fig~\ref{fig:48columnar}), predicting a phase boundary which initially scales as $T^+_c \propto \Gamma^2$.
The phase diagram for the isotropic model is shown in Fig~\ref{fig:phaseequal}. 

\begin{figure}
    	\centering
    	\includegraphics{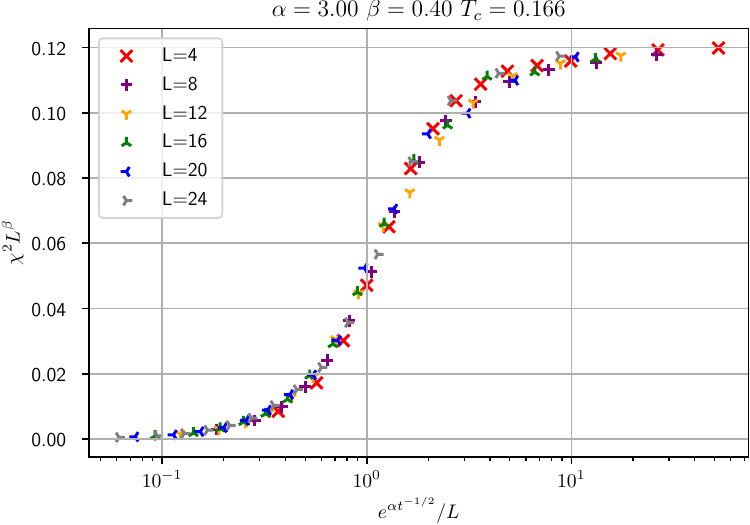}
    	\caption{Data collapse for susceptibility of $\phi$ near the thermal phase boundary between the KT phase and the paramagnet. $2\Gamma=J=J'$, with $t = \frac{T - T_c^+}{T_c^+}$. Note $T$ decreases from left to right towards $T_c^+$ which lies at  $+\infty$ on these axes. See Sec. \ref{sec:qmc} for simulation details.}
    	\label{fig:ktz8paramagnettransitiontemp}
\end{figure}

We confirm the qualitative $\Gamma>0$ phase diagram using QMC-SSE (see Sec. \ref{sec:qmc}).
First we confirm the presence of a KT phase below $T_c^+$ by analyzing sampled spin configurations at $\Gamma = J/2 < \Gamma_c$ and varying $\beta = 1/T$ and lattice sizes.
The susceptibility of $\phi$ collapses with a universal scaling function\cite{King_2018}.
This data collapse indicates $T^+_c \approx 0.166J$ for $\Gamma = \frac{J}{2}$, see Fig~\ref{fig:ktz8paramagnettransitiontemp}.

\begin{figure}
    	\centering
    	\includegraphics{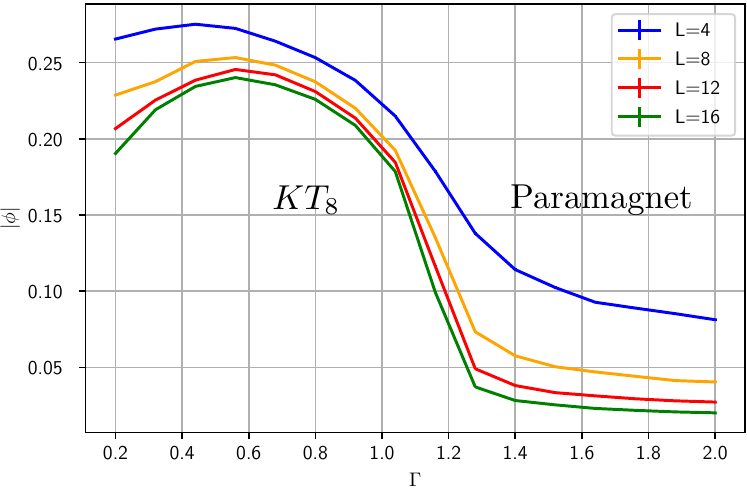}
    	\caption{Average magnitude $\vert\phi\vert$ at $\beta J = 10$ for various $L \times L$ isotropic lattice sizes showing crossover into a paramagnet near $\Gamma_c \sim 1.1J$. Error bars obscured by trendlines.}
    	\label{fig:ktz8paramagnettransition}
\end{figure}

The lower temperature phases and boundaries are more challenging to simulate.
By taking line cuts varying $\Gamma$ at $T = J/10 < T^+_c$ we map the paramagnetic transition out of the KT phase, Fig~\ref{fig:ktz8paramagnettransition}.
This finite temperature transition is expected to occur below the $\Gamma_c$ predicted by the single mode approximation, and indeed we find a transition around $\Gamma \approx 1.1J$.
Detailed scaling is inaccessible due to the slowing of the simulation at low temperatures.

The transition to the ordered $\ZZ_8$ phase lies at a temperature well below the upper critical temperature.
Actually observing the $\ZZ_8$ order phase at low temperature is numerically challenging as we expect\cite{Jose_Kadanoff_Kirkpatrick_Nelson_1977} $T_c^- \sim T_c^+/8^2$.
Furthermore competition between columnar and plaquette ordering patterns make this phase transition  particularly difficult to study, as has been observed with analogous states on the square lattice\cite{Yan_Zhou_Syljuasen_Zhang_Yuan_Lou_Chen_2021}.

\subsection{Strongly Anisotropic Model}
\label{sec:stronglyanisotropicmodel}

There are two anisotropic limits to consider: $J' \ll J$ and $J' \gg J$. 
For both we start our analysis from the classical ground state.
The $J' \ll J$ region is comparatively simple because the ordered \emph{staggered} phase persists until a critical $\Gamma_c$ or $T_c$, after which the model transitions to a paramagnet.
In the $J' \gg J$ limit we find a story similar to the isotropic case; a classical $T=0$ dimer liquid, now on a honeycomb lattice, has its ground state degeneracy lifted by a transverse field. 
Prior work on the triangular fully frustrated Ising model, and SMA analysis on the effective brick lattice (Fig~\ref{fig:48embeddingtriangular}) both suggest a $\ZZ_6$ symmetry breaking ordering pattern which can be captured by the complex order parameter,

\begin{align}
    \psi &= \sum_{R, \mu} b_\mu e^{i k R} \sigma^z_\mu \\
    b &= \frac{1}{2}(e^{\frac{2i\pi}{3}}, e^{\frac{2i\pi}{3}}, e^{-\frac{2i\pi}{3}}, 1)
    \label{eq:orderparamsymm6}
\end{align}

By analysing the action of $\mathcal{G}'$ on $(\psi, \psi^\dagger)$; we find a 6-fold clock symmetry is generated by $G_2 T_x : \psi \rightarrow e^{-\frac{i \pi}{3}} \psi$.
$T_x^2 M_{xy}: \psi \rightarrow \psi^\dagger$ acts as conjugation.
The corresponding Landau free energy admits anisotropy at $6^{th}$ order,

\begin{align}
    f' &\simeq \sum_{n=1}^{3} c'_n \vert \psi \vert^{2n} + g_6 \vert \psi \vert^6 \mbox{cos}(6\theta).
    \label{eq:landauf6}
\end{align}

Again, the anisotropy is irrelevant at the 3D XY quantum critical point. At finite temperature we expect a KT phase between upper and lower critical temperatures.
The phase diagram is very similar to that of the isotropic model (Fig~\ref{fig:phaseequal}), albeit with a critical temperature which initially scales as $\Gamma^4$.
This regime has been studied before\cite{King_2018} for a specific value of $J' = 1.8J$.

\subsection{Small Anisotropy Phase Competition}
\label{sec:weaklyanisotropicmodel}

For small anisotropy, where $\vert\frac{J'}{J} - 1\vert \ll 1$, energy corrections explicitly favor a subset of the $\ZZ_8$ ground states: columnar phases aligned with one diagonal or the other. 
This results in two distinct $\ZZ_4$ symmetry breaking columnar phases on either side of the isotropy line.
At larger anisotropy, where $J'/J$ is far from $1$, the columnar phases should give way to $\ZZ_2$ or $\ZZ_6$ phases as discussed previously.

Let us estimate the boundary separating these phases by comparing the variational energy calculated for each.
First consider the phase boundaries in the small $\Gamma$ limit, near the classical line.
Small $J'$ and $\Gamma$ perturbations about the isotropic classical point, written $\epsilon = \frac{J' - J}{J}$ and $\gamma = \frac{\Gamma}{J}$, modify $H$ by a small term $V$:

\begin{align}
    H &= J \sum_{\langle ij \rangle} s_{ij} \sigma_i^z \sigma_j^z + V \\
    V/J &= \epsilon \sum_{\substack{\langle ij \rangle \\ s_{ij} = -1}} \sigma_i^z \sigma_j^z + \gamma \sum_i \sigma_i^x
\end{align}

\noindent
We find in the $\epsilon > 0$ region, the competition is between $\ZZ_4$ states and $\ZZ_6$ states.
First, we consider the variational energy of the columnar state like the one in Fig~\ref{fig:48columnar}, labeled as $\ket{\flippablehorz}$.
Nearby low energy states can be reached by flipping two spins in any of the flippable plaquettes.
We label the state made by flipping the plaquette at unit cell $R$ as $\ket{\flippablevert}_R$.
There are as many of these states as there are unit cells, so the Hamiltonian in the dimer language\cite{Rokhsar_Kivelson_1988} restricted to these states is written

\begin{equation}
    H/J = \sum_R \epsilon \ket{\flippablevert}_R\bra{\flippablevert}_R - \gamma^2 \left(\ket{\flippablevert}_R\bra{\flippablehorz} + h.c.\right).
\end{equation}

For the general Hamiltonian there are also terms linear in $\gamma$ from flipping single spins, such as $\gamma \ket{\tripledimerdown}_R\bra{\flippablehorz}$, however they make identical contributions to all classical ground states and can be omitted when comparing the states.
At $\epsilon=0$ and $\gamma>0$ the state is independent of $\gamma$, the energy per unit cell is simply $\langle \mathcal{H} \rangle \propto -\gamma^2 J$.
For $\epsilon>0$, with finite $\gamma$, we find $\langle \mathcal{H} \rangle \propto \epsilon J$ as resonant plaquettes break the strengthened bonds.
To leading order in each expansion parameter we find the variational energy per unit cell to be $\langle \mathcal{H} \rangle =(\epsilon - \gamma^2)J$.

Repeating a similar procedure for the $\ZZ_6$ phase we need to use flippable plaquettes on the brick lattice, made by rotating 3 dimers along the alternating walls of the bricks. 

\begin{equation}
    H/J = -\sum_R \gamma^4 \left(\ket{\flippablehexa}\bra{\flippablehexb} + h.c.\right)
\end{equation}

\begin{figure}
	\includegraphics{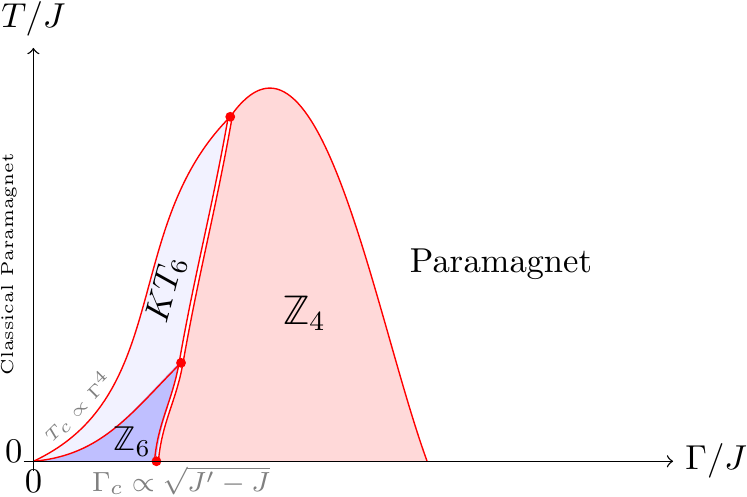}
	\caption{Temperature-transverse field phase diagram for weak anisotropy $\frac{J'-J}{J}\ll 1$. The $\psi$ order parameter with $\ZZ_6$ anisotropy permits an intermediate temperature KT phase. No such phase exists for $\phi$ with $\ZZ_4$. Double lines indicate first order transitions between these symmetry incommensurate phases; alternatively, it is possible that the phase transition splits into a region of incommensurate phases.}
	\label{fig:phasez6z4}
\end{figure}

\noindent
Since the bricks do not include dimers which break $J'$ bonds we end up with an energy per unit cell simply modified by $-\gamma^4 J$. 
Competition between these two energies determines the shape of the boundary between the $\ZZ_4$ and $\ZZ_6$ phases at low $\epsilon$ and low $\gamma$. 
To leading order this is given by $\epsilon_c \propto \gamma^2$.
In the $J' < J$ region the staggered phase competes with a columnar phase (rotated 90 degrees from that shown in Fig~\ref{fig:48columnar} so as to break a single $J'$ bond per unit cell). 
The same treatment predicts a similar phase boundary below the $J' = J$ line: $\epsilon_c \propto -\gamma^2$.
These estimates also motivate the $T_c$ boundaries shown in Fig~\ref{fig:phaseequal} and Fig~\ref{fig:phasez6z4}.

Notably KT phases are not supported by Clock Models with $q \le 4$ states\cite{Jose_Kadanoff_Kirkpatrick_Nelson_1977}. 
Despite the isotropic and strongly anisotropic models having KT phases, some of the phases of the weakly anisotropic models, such as the $\ZZ_4$ phase, do not support KT phases and instead transition directly into the thermal paramagnet above a critical temperature.

\subsection{Incommensurate Phases}
\label{sec:incommensuratephases}

\begin{figure}
	\centering
	\includegraphics{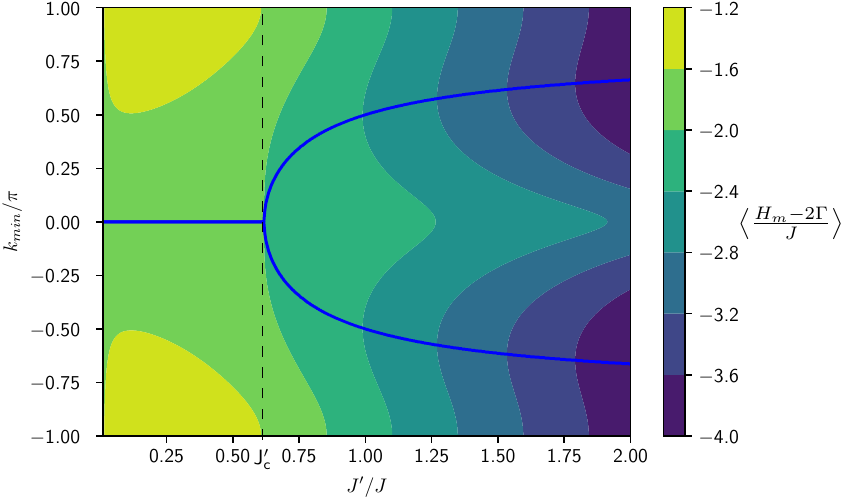}
	\caption{Evolution of energy of lowest magnon with momentum $\vec{k}=(k,-k)$ with anisotropy $J'/J$. Minimum momentum configurations for each value of $J'/J$ are traced.}
	\label{fig:minimummomentumplot}
\end{figure}

Between the weakly anisotropic and strongly anisotropic limits previously described, we find a collection of additional ground state ordering patterns with distinct momenta.
SMA in the large $\Gamma$ limit predicts a smoothly varying magnon condensation momentum $\vec{k}_{min}=\pm(k_{min},-k_{min})$ as a function of $J'$, see Fig~\ref{fig:minimummomentumplot}.
We can trace $k_{min}$ as we increase $J'$.
As seen previously, $J' = J$ predicts $k_{min} = \frac{2\pi}{4}$.
Increasing to $J'\approx 2 J$ moves $k_{min}$ towards $\frac{2\pi}{3}$.
Between these values, however, SMA predicts ordering patterns which are incommensurate with the underlying lattice\cite{Bak_1982}.
We expect the the model to lock into ground states with smaller unit cells, in other words smaller denominators $q$ in $k_{min}=\frac{p\pi}{q}$.
In the $J' \gg J$ limit SMA predicts $k_{min} \sim \frac{4 \pi}{5}$.
This disagrees with the SMA prediction on the large $J'$ effective brick lattice (Fig~\ref{fig:48embeddingtriangular}), as well as the perturbative expansion around the classical $\Gamma=0$ line, so we expect no $k_{min} = \frac{4 \pi}{5}$ phase to appear in the phase diagram.

\begin{figure}
	\centering
	\includegraphics{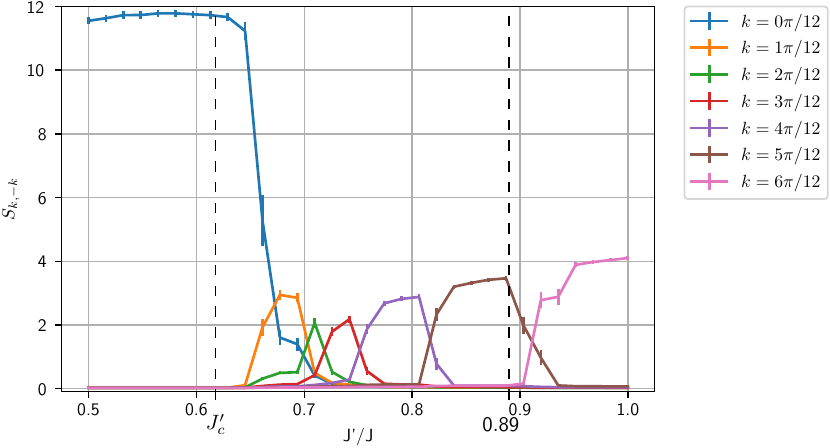}
	\caption{Structure factor at momentum $(k,-k)$ at varying $J'$. Samples at $\beta J = 10$ and $\Gamma = 0.9J$. Results of QMC study on 24x24 lattice. Vertical dashed lines indicate illustrative transitions predicted by the SMA (left, bifurcation of minimum $k$ in magnon dispersion, cf. Fig.~\ref{fig:minimummomentumplot}; right, crossing of energy at $k=2\pi/4$ and $k = 2\pi/5$).}
	\label{fig:z4staggeredstaircase}
\end{figure}

\begin{figure}
    \centering
	\includegraphics{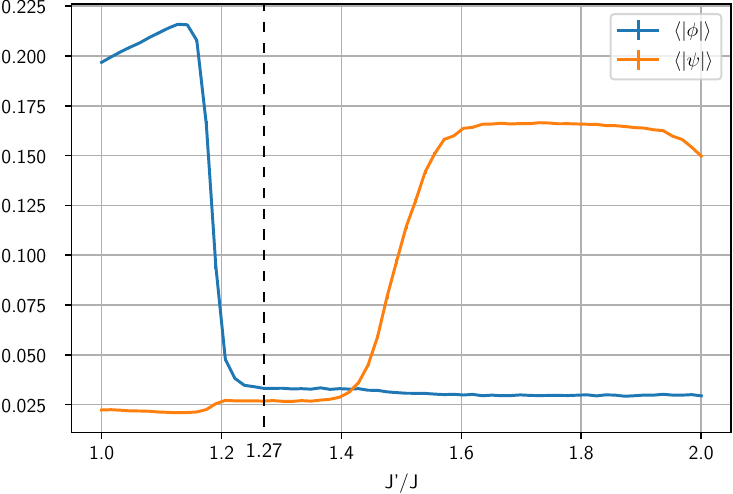}
	\caption{Transition from $\ZZ_4$ phase to a quasi-long range ordered $\ZZ_6$ KT phase. QMC data from a 12x12 lattice at $\beta J = 10$ and $\Gamma = J$. 
	Vertical dashed line indicates energy crossing of $k=2\pi/3$ and $k=2\pi/4$ in SMA dispersion. Error bars obscured by trendlines.}
	\label{fig:z4toz6highgammatransition}
\end{figure}

We can predict the approximate phase boundaries by numerically comparing the eigenenergies of magnon excitations.
Expanding $H_m$ (\ref{eq:hmagnon}) around $k_{min}=0$ (which corresponds to the \emph{staggered} phase) predicts a critical value of $J_c'= \left(\frac{1 + \sqrt{5}}{3 + \sqrt{5}}\right)J$ at which the zero mode is no longer the first to condense.
Magnons with momentum $k = \frac{2\pi}{5}$ and those with $k = \frac{2\pi}{4}$ have equal energy at $J' \approx 0.89 J$.
For large enough $\Gamma$ we therefore expect a first-order phase transition near $J' \approx 0.89 J$ out of a phase with $k = \frac{2\pi}{5}$, and another into a $\ZZ_4$ symmetry breaking phase with $k = \frac{2\pi}{4}$ (Fig~\ref{fig:z4staggeredstaircase}). 
There may, however, be additional symmetry breaking phases between these.
Similarly we predict a transition between the upper $\ZZ_4$ breaking phase and the $\ZZ_6$ breaking phase (with $k = \frac{2 \pi}{3}$) to occur near $J' \approx 1.27 J$ (Fig \ref{fig:z4toz6highgammatransition}).
Note that these predictions are from SMA and performed in the infinite $\Gamma$ limit, far above the transition into the paramagnet. 
Despite this, in both Fig~\ref{fig:z4staggeredstaircase} and Fig~\ref{fig:z4toz6highgammatransition} the predicted phase boundaries line up very well with observed transitions from QMC samples.

\section{Quantum Monte Carlo}
\label{sec:qmc}

We confirm many features of this complicated phase diagram using QMC-SSE\cite{Sandvik_Kurkijarvi_1991,Sandvik_2003,Sandvik_2019,Melko_2013} to sample $z$-aligned spin configurations $\ket{\psi}$ from the lattice according to their Boltzmann weights, $\bra{\psi}e^{-\beta H}\ket{\psi}$.
These samples can then be used to compute observables.
SSE requires all terms in the Hamiltonian to be positive; we use the transverse field Ising model Hamiltonian plus a constant offset restricted to $\Gamma \ge 0$:
\begin{equation}
    H = \sum_{\langle ij\rangle} J_{ij} (\sigma_i^z \sigma_j^z + I) - \sum_{i} \Gamma (\sigma_i^x + I)
\end{equation}

We use typical diagonal and cluster updates, as well as a custom semiclassical update for improving performance near the $\Gamma = 0$ limit (Sec. \ref{section:rvb}).
This update exploits flippable plaquettes more readily than cluster updates by moving diagonal terms between bonds.
We also use Replica~swapping~/~parallel~tempering\cite{Hukushima_Nemoto_1996} for sweeps across parameters $\beta$, $\Gamma$, and $J'$.

\begin{figure*}
    \begin{subfigure}[b]{0.32\textwidth}
        \centering
    \includegraphics[width=\linewidth]{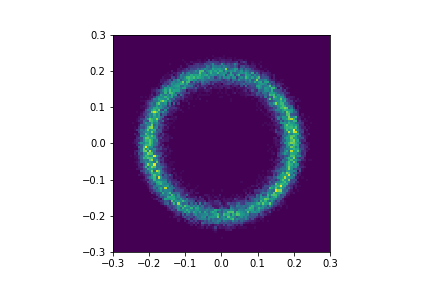}
	\caption{$\Gamma = J$ and $\beta J = 10$}
	\label{fig:phi8ktplanelowT}
    \end{subfigure}
    \begin{subfigure}[b]{0.32\textwidth}
        \centering
	\includegraphics[width=\linewidth]{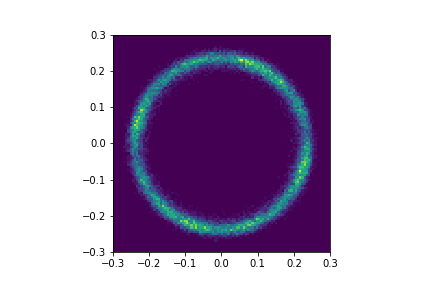}
	\caption{$\Gamma = 0.8J$ and $\beta J = 256$}
	\label{fig:phi8columnarplanelowT}
    \end{subfigure}
    \begin{subfigure}[b]{0.32\textwidth}
    	\centering
    	\includegraphics[width=\linewidth]{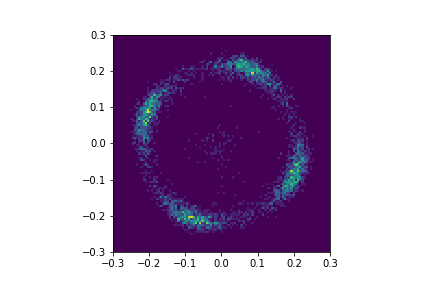}
    	\caption{$J' = 1.1J$, $\Gamma=J$, $\beta J = 10$}
    	\label{fig:phi8columnarplane}
    \end{subfigure}
    
    \caption{ Probability density of $\phi$ sampled in complex plane from QMC-SSE. Note the bulls eye structure indicates that the amplitude has developed an expectation value even if the phase continues to fluctuate as in the (a) KT phase. At lower $T$, (b) shows subtle evidence of $\ZZ_8$ symmetry breaking corresponding to a columnar phase. With anisotropy, (c) shows 4-fold symmetry breaking corresponding to columnar order.
    }
\end{figure*}

To study phase transitions we take the sampled $z$-basis spin configurations and act on them with the relevant complex order parameter operators (e.g. Eq~(\ref{eq:orderparamsymm}),~(\ref{eq:orderparamsymm6})).
Within the appropriate phase the order parameters acquire a non-zero vacuum expectation value; the Monte-Carlo sampling density supports this Fig~\ref{fig:phi8ktplanelowT}-\ref{fig:phi8columnarplane}.

To find phase transitions we look at averages of the magnetization $m  = \sqrt{\phi^* \phi}$, or susceptibility $\chi = m^2$.
Each of these averages is across 4 independent simulations and 1000 samples for each, samples are taken after 100 or 1000 QMC sweeps (depending on parameters), beginning after a 100,000 sweep thermalization period.
These sampling rates were determined by measuring the autocorrelation time of the bond variables in the $KT_8$ phase.
A sweep consists of diagonal update, a cluster update, and a thermal spin-flip update on any variables unconstrained by imaginary time operators. 
At low $\Gamma$ we additionally perform $\frac{N}{2}$ semiclassical updates with $N$ the total number of spins.
These do not make a significant difference to autocorrelation times for $\Gamma > T$ and can be omitted for such simulations.

\subsection{Semiclassical RVB Update}
\label{section:rvb}

Multibranch cluster updates\cite{Sandvik_2003,Melko_Sandvik_2005} perform poorly as the model approaches the classical low $\Gamma$ regime and the entire SSE graph forms a single cluster connected by two-body interactions.
To improve autocorrelation times for the simulation near the classical limit we have developed an update which flips smaller clusters of spins, motivated by the classical ring exchange updates for dimer models.

Consider an Ising-symmetric model with diagonal 2-body operators ($H^d_{ij} = J_{ij} \sigma_i^z \sigma_j^z$) and offdiagonal 1-body operators ($H^{t}_i = \Gamma \sigma^x_i$).
We can make $H^{t}$ constant across all matrix elements, $[H^{t}_i]_{ss'} = -\Gamma$, and ensure no diagonal matrix elements are negative by adding a constant to the Hamiltonian

\begin{equation}
    H = (J_{ij}\sigma_i^z \sigma_j^z + JI) - \Gamma (\sigma_x + I).
\end{equation}

The RVB update consists of flipping clusters spanning real space and imaginary time, similarly to the multibranch cluster update.
This update differs in that we allow 2-body diagonal operators to be moved around on the surface of the cluster. 
The multibranch update can then be viewed as the special case where there are no operators on the surface, and therefore no operators need to be moved.

\begin{figure}
	\centering
	\includegraphics{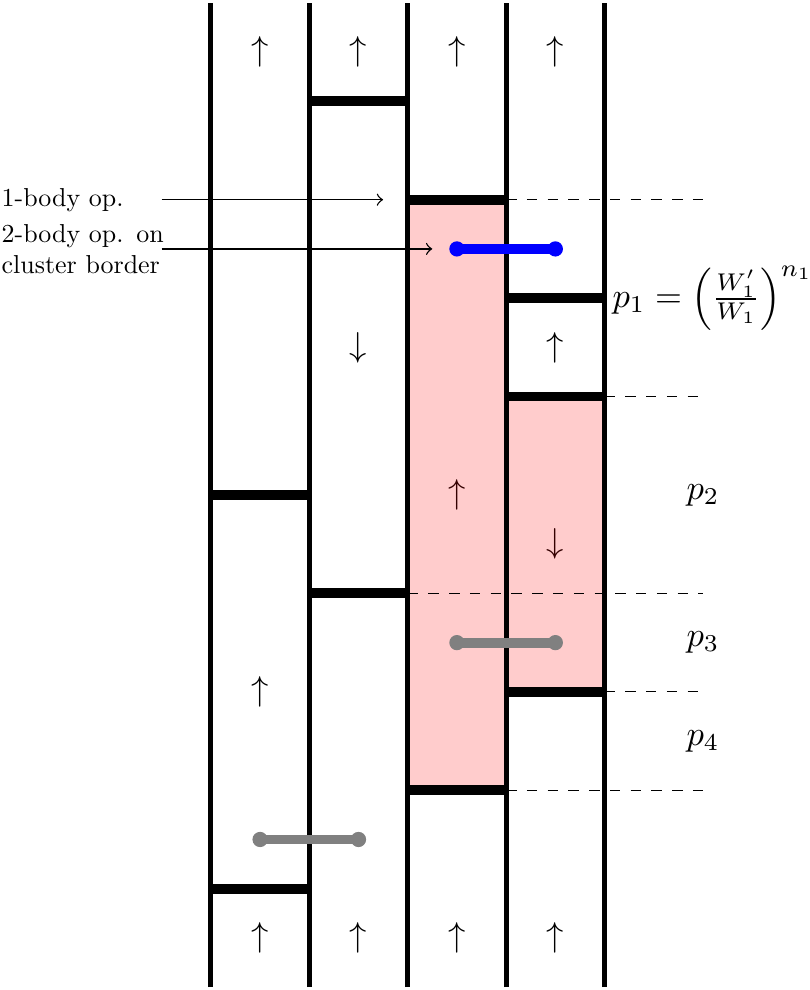}
	\caption{An example cluster for the semi-classical RVB update. The four relevant segments are separated by dashed lines. Spin world lines are represented by the white space between vertical black lines. The red shaded cluster will be flipped and the blue two-body operator moved by the update if accepted.}
	\label{fig:rvbupdate}
\end{figure}

To visualize this update, first consider world lines for each spin (Represented in Fig~\ref{fig:rvbupdate} as the white space between vertical black lines), and boundaries (horizontal lines) defined by the constant 1-body operators.
Since the 1-body operators have constant matrix elements across all indices, when flipping a world line we may stop at any of the 1-body operators without changing the weight of the graph.
Define a cluster by selecting subsections of the world lines (highlighted in red).
Each subsection must begin and end at one of the 1-body operators, like the multibranch cluster update.
A convenient definition of the cluster is the set of spins within it at each point in imaginary time: $\mathcal{C}(\tau)$.

To flip the spins in the cluster we will move the 2-body operators with one leg inside the cluster, and one leg outside.
These are the operators whose weights may be changed by the new spin configuration. 
To simplify the update, for each operator to be moved at imaginary time $\tau$, we will only consider new positions at the same $\tau$.
At each moment in imaginary time we will calculate the acceptance probability of moving a 2-body operator on the border, $p(\tau)$.
We can group contiguous regions of imaginary time with identical classical spin configurations in or around the cluster into \emph{segments} with the same $p(\tau)$.
The total flip probability for segment $\alpha$ is then $p_\alpha = p(\tau)^{n_\alpha}$ where $n_\alpha$ the number of operators needing to be moved in that segment.
Segment delimiters are marked as dashed lines in Fig~\ref{fig:rvbupdate}.

Let $\ket{\psi_\alpha}$ be the classical spin state in and around the cluster in segment $\alpha$.
Let $\ket{\psi'_\alpha}$ be the similarly defined spin state after the cluster has been flipped.
Define $W_\alpha = \sum_{\langle i/j \rangle} \bra{\psi_\alpha} H^d_{ij} \ket{\psi_\alpha}$ the energy contribution from diagonal terms along the border of the cluster, and similarly $W'_\alpha$ for $\psi'_\alpha$.
\begin{align}
    \langle i / j \rangle \equiv \{ i, j \in \langle ij \rangle ~\vert~ \left(i \in \mathcal{C}(\tau) \right) \oplus \left(j \in \mathcal{C}(\tau)  \right) \}
\end{align}
\noindent
The notation $\langle i/j \rangle$ indicates nearest neighbors $i$ and $j$ where exactly one of $i$ or $j$ is inside the cluster at a given imaginary time slice.

For an update rule from state $a$ to $b$, detailed balance requires $\mathbf{P}_{accept} \frac{\mathbf{P}_{a \rightarrow b}}{\mathbf{P}_{b \rightarrow a}} = \frac{W'}{W}$ with $W$ the total Boltzmann weight of the graph. 
For each segment there are $n_\alpha$ operators to move, each one represents a term in $H^d$, let 
\begin{align}
    W_\alpha &= \sum_{\langle i / j \rangle} \bra{\psi_\alpha} H^d_{ij} \ket{\psi_\alpha} = \sum_l w_{\alpha,l} \\
    W_\alpha' &= \sum_l w'_{\alpha,l}
\end{align}
\noindent
where each $w_l$ represents one of the sites along the border of the cluster, and an entry in $\langle i/j \rangle$.
The proposed state $b$ is made by flipping the spins in the cluster then moving all the $n_\alpha$ terms on the border to new positions $l'$ with probability $\frac{w'_{l'}}{W'_\alpha}$.
Let $\mathbf{P}_{a\rightarrow b}$ be the probability of selecting a specific state $b$ after the cluster has flipped from state $a$,
\begin{align}
    \mathbf{P}_{a\rightarrow b} &= \prod_\alpha \prod_{i=1}^{n_\alpha} \frac{w'_{\alpha,b_i}}{W'_\alpha} \\
    \mathbf{P}_{b\rightarrow a} &= \prod_\alpha \prod_{i=1}^{n_\alpha} \frac{w_{\alpha,a_i}}{W_\alpha}
\end{align}
\noindent
where $a_i$ and $b_i$ are used to indicate the new positions for each of the $n_\alpha$ operators to make state $a$ and $b$.

We can now define the acceptance weight for segment $\alpha$ as 
\begin{align}
    p_\alpha = \left( \frac{W'_\alpha}{W_\alpha}\right)^{n_\alpha}
\end{align}
and show that it satisfies detailed balance with a metropolis style update rule.
An operator in segment $\alpha$ which moves from position $a_i$ to $b_i$ has a relative weight change $\frac{w'_{\alpha,b_i}}{w_{\alpha,a_i}}$, implying the net graph weight change is $\frac{W'}{W} = \prod_{\alpha}\prod_i^{n_\alpha} \left(\frac{w'_{\alpha,b'_i}}{w_{\alpha,a_i}}\right)$.
We can now solve for the probability of accepting the update:

\begin{align}
    \mathbf{P}_{accept} &= \frac{W'}{W} \frac{\mathbf{P}_{b \rightarrow a}}{\mathbf{P}_{a\rightarrow b}} \nonumber \\
    &= \left(\prod_{\alpha,i} \frac{w'_{\alpha, b_i}}{w_{\alpha, a_i}}\right) \left(\frac{\prod_{\alpha,i} \frac{w_{\alpha,b_i}}{W_\alpha}}{\prod_{\alpha,i} \frac{w'_{\alpha,b_i}}{W'_\alpha}}\right) \nonumber \\ 
    &= \prod_{\alpha,i} \left(\frac{W'_\alpha}{W_\alpha} \right) = \prod_\alpha p_\alpha
\end{align}

\noindent
As is standard in metropolis updates, $\mathbf{P}_{accept}$ is taken as the minimum of this calculated value and $1$.
By choosing clusters to update based on the flippable plaquettes in the expected ground state, we increase the likelihood of having segments with classically flippable clusters, and thus $p_\alpha = 1$.

By randomly selecting small candidate clusters we were able to improve the autocorrelation times in the small $\Gamma$ regime (See Appendix.~\ref{sec:rvbautocorr}).
This update move is limited by the choice of cluster selection,
and we have yet to investigate applying other cluster based update moves to the cluster selection stage, though some candidates exist such as the sweeping cluster update\cite{Yan_2019}.
Furthermore, although the RVB update was motivated by classical ring exchange updates, there may be room for improvement by adapting other algorithms developed for classical clock models\cite{tomita2002} in a similar way.

\section{Realization on a Quantum Annealer}
\label{sec:annealer}

To experimentally test some of the predicted phases we use a programmable quantum device.
Initially built to study classical optimization problems, the \mbox{D-Wave 2000Q} is a superconducting qubit based quantum annealer which simulates a TFIM Hamiltonian.

\begin{equation}
    H = A(s) \sum_i \sigma^{x}_i + B(s) \left( \sum_{\langle ij \rangle} J_{ij} \sigma_i^{z} \sigma_j^{z} + h_i \sigma_i^z \right)
    \label{eq:tfimdwave}
\end{equation}

\begin{figure}
	\centering
	\includegraphics[width=0.8\linewidth]{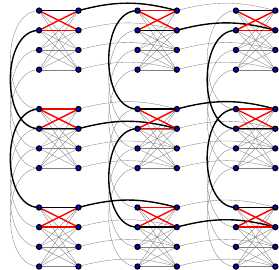}
	\caption{Subsection of D-Wave's Chimera Architecture. Spin variables live on vertices with tunable Ising interactions shown as edges. Highlighted edges (identical color scheme to Fig~\ref{fig:48embedding}) correspond the embedding of the 4-8 lattice with open boundary conditions. All other couplings disabled.}
	\label{fig:chimera}
\end{figure}

\noindent
Here, $J_{ij}$ and $h_i$ are programmable parameters, and $s$ is the annealing variable which tunes the relative strengths of $A(s)$ and $B(s)$.
The default annealing schedule takes $s$ from $0$ to $1$, starting at $B(0) = 0$ and ending at $A(1) = 0$, each following a curve specific to the machine\cite{D-WaveDocs}. 
The set of available couplings is given by the architecture; the Chimera architecture of the \mbox{D-Wave 2000Q} provides up to 2048 qubits and 4196 tunable couplings arranged in a 16x16 grid with unit cells of 8 qubits each, a subsection is shown in Fig~\ref{fig:chimera}.

\begin{figure}[h]
    \centering  
    \includegraphics[width=246.0pt]{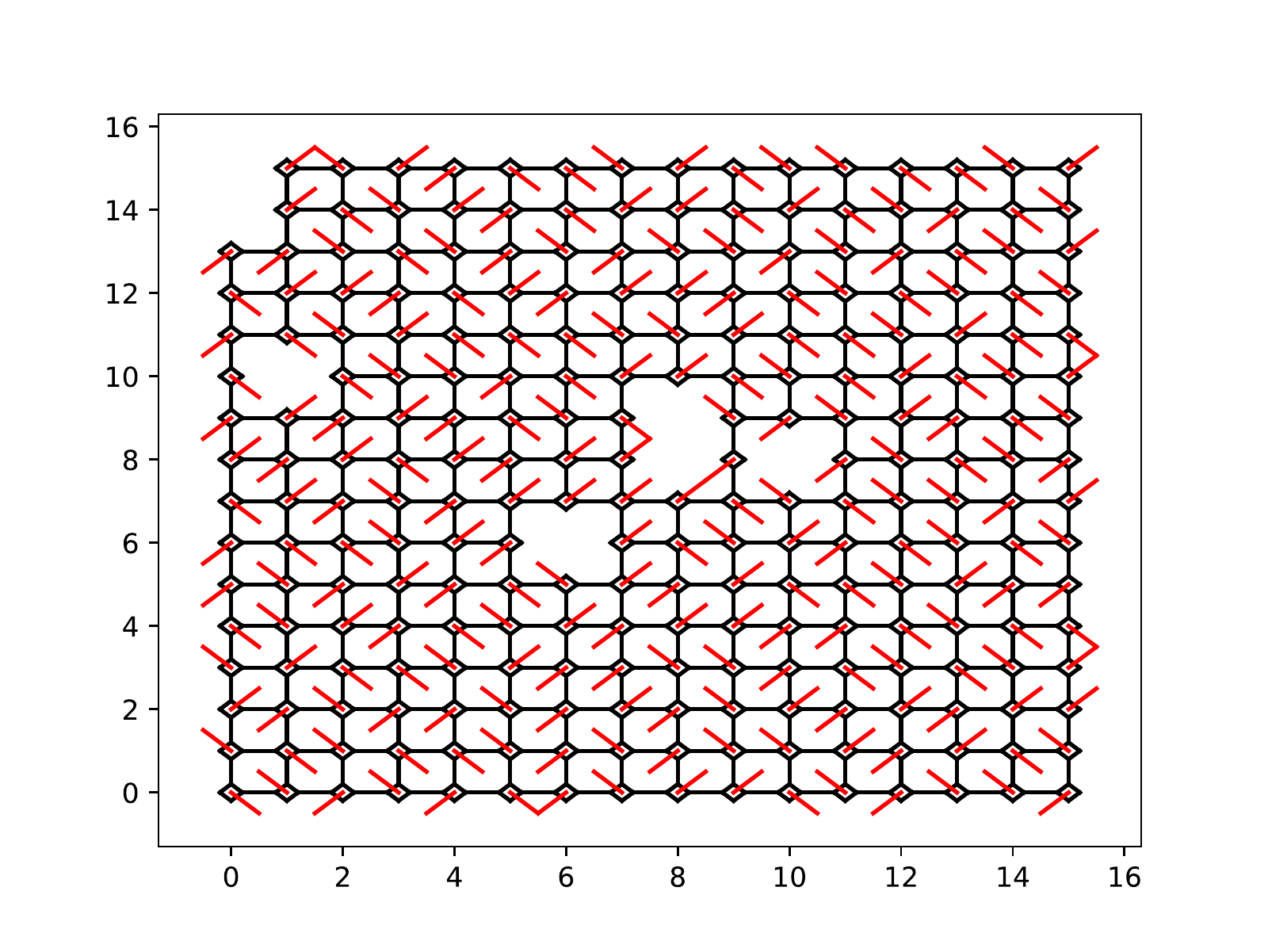}
    \caption{Samples from the D-Wave 2000Q can be profitably visualized through the dimer mapping.}
    \label{fig:dwavedimersample}
\end{figure}

To corroborate the evidence provided by Quantum Monte Carlo for the predicted phase diagram, we embedded the 4-8 lattice isotropic Ising model in the Chimera architecture. 
Some defects are introduced by missing spin variables or bonds.
These are machine-dependent defects and can not be completely avoided, see Fig~\ref{fig:dwavedimersample} for an example embedding and sampled dimer state.
Each unit cell in the Chimera graph contains 8 spin variables, allowing two separate sheets of the 4-8 lattice to be embedded together then tied at their borders. 
Both cylindrical\cite{King_2018} and open boundary conditions can be embedded.

\begin{figure}[h]
        \centering
        \includegraphics{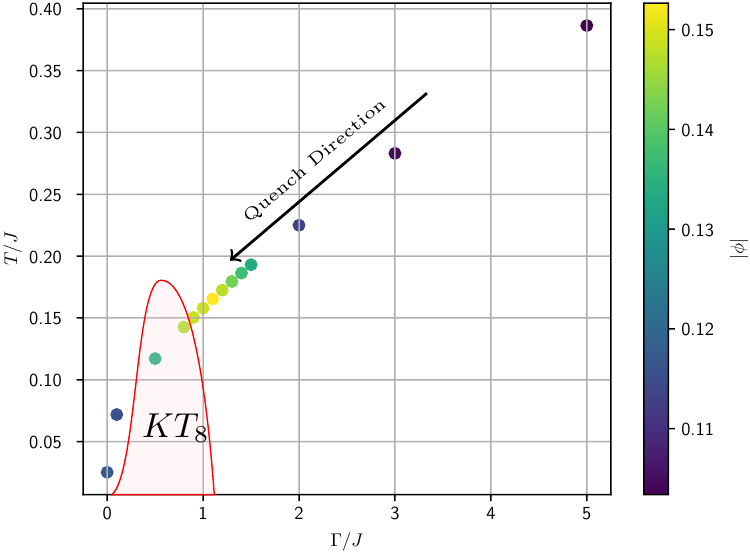}
       \caption{Trajectory through the $T/\Gamma$ plane. $\vert\phi\vert$ shown for Cylindrical b.c. The phase boundary is drawn consistent with theory (at small $\Gamma$) and QMC line cuts in Fig~\ref{fig:phaseequal};  we have not systematically mapped out the entire boundary.}
        \label{fig:dwavephase}
\end{figure}

We can alter the default annealing schedule to allow the system to equilibrate at nonzero transverse field.
Rather than scaling directly from $s = 0$ to $1$, we scale $s$ to chosen value $s_\circ$ such that $\Gamma / \vert J \vert = A(s_\circ) / B(s_\circ)$ and allow the system to thermalize.
The machine has a low physical temperature, $T \sim 14mK$, and relatively large coupling strengths $\beta B(s=1) \sim 41$ (and $\beta A(s=0) \sim 34$). 
However the couplings vary as a function of the annealing schedule $s$, therefore a limitation of this approach is that since both $A$ and $B$ change as a function of $s$, we find a lower bound for $T / \vert J \vert$ for a given choice of $\Gamma / \vert J \vert$, producing an annealing path which varies in both $\Gamma/\vert J \vert$ and $T/\vert J \vert$ (Fig~\ref{fig:dwavephase}).
Furthermore the \mbox{D-Wave 2000Q} must bring the transverse field to $\Gamma=0$ before taking spin measurements.
The timescale of this quench is not negligible compared to the inverse energy scales, and may be a source of noise.
An additional machine limitation is the precision of the $J_{ij}$ and $h_i$ couplings, which may be shifted from their expected values.
These shifts may be addressed by shimming\cite{King_2018} -- performing random measurements and using average magnetization and correlations to modify couplings.
Previous work\cite{King_2018, King_2021} has additionally included more advanced noise reduction techniques, such as adjusting bonds based on symmetries of the model, altering annealing schedules, and performing reversed or repeated annealings from seeded ordered states.

\begin{figure}
        \centering
        \includegraphics{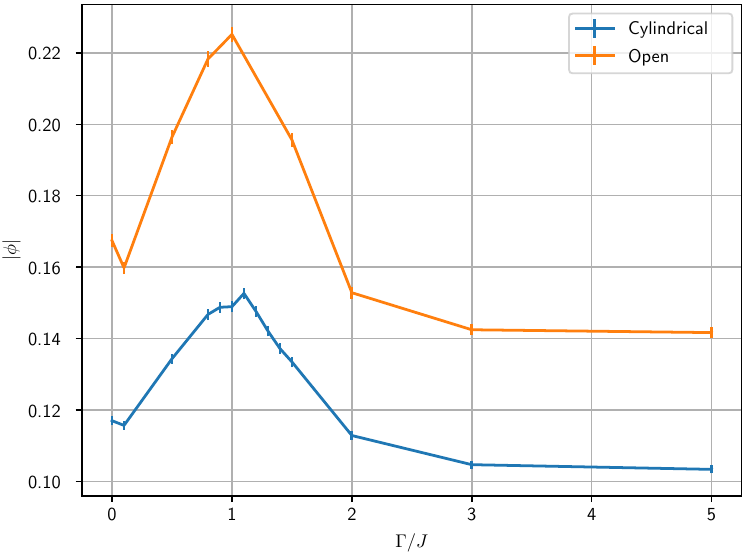}
        \caption{Evidence of $\ZZ_8$ KT quasi-long range ordering in D-Wave. Data from 2048 samples for each point. Cylindrical b.c. used 1866 qubits on two sheets, Open b.c. used 996 qubits on a single sheet.}
        \label{fig:dwaveabsorder}
\end{figure}

By sampling at the lowest allowed $T/\vert J \vert$ for each choice of $\Gamma$, or in other words using the largest allowed coupling strengths, we find a significant increase in $\vert \phi \vert$ around the expected transverse fields. 
Since larger $\Gamma$ values must be quenched and pass through phase boundaries we expect the response to be smeared toward larger $\Gamma$ values.
The result are summarized in Fig~\ref{fig:dwaveabsorder}-\ref{fig:dwavephase}.
Qualitatively, we find $\phi$ to be consistent with the predictions and numerics in preceding sections, although the data quality falls far below that which QMC is able to offer.

\section{Conclusion}

The fully frustrated 4-8 lattice contains a deceptively large variety of 2D phases of matter.
The phase diagram is developed in low and large $\Gamma$ limits for both the isotropic and anisotropic models, and numerical methods support our theoretical arguments. 
The \mbox{D-Wave 2000Q} quantum annealer is nominally able to realize the frustrated lattice, and by altering the annealing schedule we could measure the ordering effects of a transverse field.
However, the support offered by the annealer was of lower quality than that provided by QMC especially as the D-Wave device was limited to relatively high temperature regions of the phase diagram at finite transverse field. 
Accessible parameters are within an order of magnitude of the couplings needed to observe the quantum ordered or KT phases and thus may well be realizable in the next generation devices.
Our study suggests that future generations of the D-Wave annealer may actually contribute to physical understanding of frustrated models beyond those accessible with current classical techniques.

\section{Acknowledgements}

The authors are grateful to C. Chamon, A. Chandran, R. Moessner, A. Sandvik, K-H. Wu, S. Zhou for useful discussions.
This research used Ising, Los Alamos National Laboratory's D-Wave quantum annealer. Ising is supported by NNSA's Advanced Simulation and Computing program. 
C.R.L. acknowledges support from the NSF through grant PHY-1752727 and the generous hospitality of the Aspen Center for Physics, which is supported by NSF grant PHY-1607611. 

\vfill

\newpage
\bibliography{mybib}

\cleardoublepage

\appendix

\section{Single Mode Approximation}
\label{sec:smaappendix}

We can find the relevant order parameters by estimating which delocalized magnon excitations in the paramagnetic region are the first to condense.
The Bloch Hamiltonian restricted to the single magnon states with momentum $k$ is given by

\begin{align}
    %\ket{\phi_k} &= \sum_{R,\mu} a_\mu e^{ik \cdot R} \sigma^z_{R,\mu} \ket{\rightarrow \rightarrow ...}\\
    H_{m} &= \left(\begin{matrix}
    0 & -J & - J e^{i k_y} & J' \\
    -J & 0 & J' & - J e^{i k_x} \\
    -J e^{-i k_y} & J' & 0 & J' \\
    J' & - J e^{-i k_x} & J' & 0
    \end{matrix}\right) + 2 \Gamma.
    \label{eq:hmagnon}
\end{align}

\noindent
The minimum eigenenergy is given by

\begin{align}
    E_k &= 2 \Gamma - J \sqrt{3 + 2 \sqrt{1 - \mbox{sin}(k_x) \mbox{sin}(k_y)}} \\
    E_{(\frac{\pi}{2}, -\frac{\pi}{2})} &= 2\Gamma - (1+\sqrt{2})J
\end{align}

\section{RVB Autocorrelation Times}
\label{sec:rvbautocorr}

\begin{figure}[h!]
	\centering
	\includegraphics[width=0.8\linewidth]{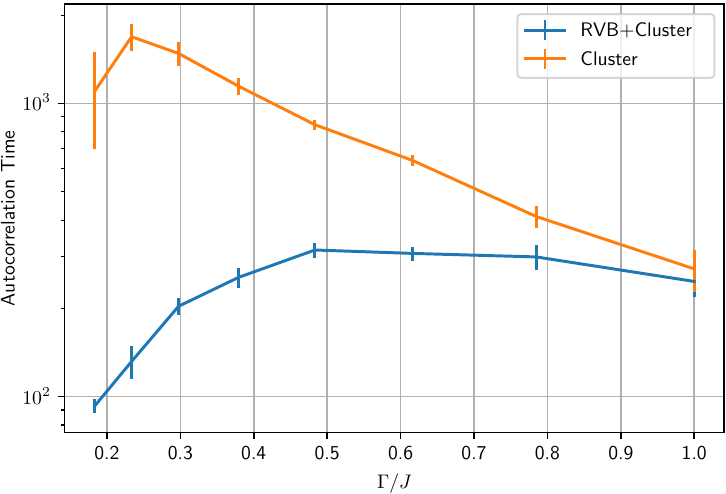}
	\caption{Autocorrelation time for bond variables with and without the RVB update step. Data taken from 32 independent runs on a $4\times4$ isotropic 4-8 lattice at $\beta J=5$.}
	\label{fig:rvbautocorrelation}
\end{figure}

To illustrate the benefit of the RVB update in the small $\Gamma$ regime we measured the exponential fit to $e^{-t/\tau}$ of the autocorrelation graphs for bond variables, sampled over a variety of transverse field values (Fig~\ref{fig:rvbautocorrelation}).
We see convergence around $\Gamma=J$ where the RVB does not significantly contribute compared to the cluster update.

\noindent

\end{document}